# Multiparametric Deep Learning and Radiomics for Tumor Grading and Treatment Response Assessment of Brain Cancer: Preliminary Results

Vishwa S. Parekh, PhD[1,2], John Laterra, MD, PhD[3,4], Chetan Bettegowda, MD, PhD[5], Alex E. Bocchieri, MSE[2], Jay J. Pillai, MD,[1, 5*] Michael A. Jacobs, PhD[1,3*]

[1]The Russell H. Morgan Department of Radiology and Radiological Sciences, T
he Johns Hopkins University School of Medicine, Baltimore, MD 21205
[2]Department of Computer Science, The Johns Hopkins University, Baltimore, MD 21208
[3]Sidney Kimmel Comprehensive Cancer Center, The Johns Hopkins University School of Medicine, Baltimore, MD 21205
[4]Department of Neurology, The Johns Hopkins University School of Medicine, Baltimore, MD 21287
[5]Department of Neurosurgery, The Johns Hopkins University School of Medicine
[5]F.M. Kirby Center for Functional Brain Imaging,  Kennedy-Krieger Institute, Baltimore, MD 21217

Address Correspondence To: Michael A. Jacobs (mikej@mri.jhu.edu) or Jay J. Pillai ( jpillai1@jhmi.edu)

*Abstract:* **Radiomics is an exciting new area of texture research for extracting quantitative and morphological characteristics of pathological tissue. However, to date, only single images have been used for texture analysis.  We have extended radiomic texture methods to use multiparametric (mp) data to get more complete information from all the images.  These mpRadiomic methods could potentially provide a platform for stratification of tumor grade as well as assessment of treatment response in brain tumors. In brain, multiparametric MRI (mpMRI) are based on contrast enhanced T1-weighted imaging (T1WI), T2WI, Fluid Attenuated Inversion Recovery (FLAIR), Diffusion Weighted Imaging (DWI) and Perfusion Weighted Imaging (PWI).  Therefore, we applied our multiparametric radiomic framework (mpRadiomic) on 24 patients with brain tumors (8 grade II and 16 grade IV).  The mpRadiomic framework classified grade IV tumors from grade II tumors with a sensitivity and specificity of 93% and 100%, respectively, with an AUC of 0.95.  For treatment response, the mpRadiomic framework classified pseudo-progression from true-progression with an AUC of 0.93. In conclusion, the mpRadiomic analysis was able to effectively capture the multiparametric brain MRI texture and could be used as potential biomarkers for distinguishing grade IV from grade II tumors as well as determining true-progression from pseudo-progression.**

*Key Terms* **- Magnetic resonance imaging, multiparametric MRI, Brain, Cancer, Diffusion, tissue biomarkers. Radiomics, multiparametric Radiomics, treatment response, progression**

## I. Introduction

Brain cancer is a major health problem in the United States. According to estimates from the American Brain Tumor Association & Central Brain Tumor Registry of the United States, there is an annual incidence of nearly 80,000 new cases of primary brain tumors in the U.S. In addition, the estimated annual mortality from primary malignant CNS tumors is approximately 17,000 [1]. Despite advances in structural and physiologic magnetic resonance imaging (MRI), positron emission tomography (PET) and other nuclear medicine imaging techniques and molecular genetics, impact on overall mortality and morbidity, particularly for high grade gliomas, remains dismal; the 5-year survival for glioblastoma patients is 15-20% despite maximal safe resection, chemotherapy with temozolomide and radiation treatment. Moreover, no therapeutic options provide substantial survival benefit for recurrent disease [2-4].

Although there have been advances in radiological imaging of human brain tumors (structural MRI, diffusion and perfusion MRI, MR spectroscopic imaging (MRS), PET, task and resting state BOLD), reliable utility has only been established in noninvasive grading of de novo tumors [5].

However, distinguishing between true progression and pseudoprogression in high grade gliomas remains a clinical challenge and is frequently accomplished only through surgical biopsy/resection and not noninvasively [6-10]. Need for better imaging assessment of progressive vs. stable disease led to the development of the recent RANO (Response Assessment in Neuro-Oncology) criteria[11-13], but even these new imaging criteria do not allow accurate and early differentiation between true therapeutic response and pseudoresponse  or between true progression and pseudoprogression; even metabolic and physiologic imaging such as PET, MR spectroscopy and amide proton transfer imaging have not been entirely reliable for such differentiation.  Furthermore, neither perfusion imaging nor MRS is 100% specific in distinguishing true progression from pseudoprogression or actual treatment response from pseudoresponse in high grade gliomas. [6-10].  The need for standardization and validation of conventional MR imaging determinants of therapeutic response has led to development of the recent Response Assessment in Neuro-Oncology (RANO) criteria [9,13]. Moreover, a recent 2016 update to the earlier 2007 WHO classification system, now includes molecular markers, such as isocitrate dehydrogenase gene (IDH1) mutation status, ATP-dependent helicase (ATRX), Telomerase reverse transcriptase (TERT), TP53, 1p/19q co-deletion, O6-methylguanine-DNA methyltransferase (MGMT) methylation status and EGFR expression in the definition of glioma subtypes based on prognostic factors and treatment responsiveness[14-18]. Furthermore, no currently available imaging biomarkers exist to reliably predict transformation to higher grades from low grade gliomas.

In recent years, radiomics has been applied to many different precision medicine and radiology applications in the research setting with great preliminary results [14,19-22]. However, radiomics was initially developed for application to single radiological images and may not characterize high dimensional datasets such as multiparametric brain MRI effectively. Recently, we developed the multiparametric radiomics (mpRadiomic) framework to extend the single radiomics framework for application to multiparametric radiological imaging setting with excellent results [23]. Here, we implement the mpRadiomic framework for application to brain mpMRI and evaluate its efficacy in brain tumor grading and treatment response assessment.

## II. METHODS AND MATERIALS

**Clinical Parameters:** Twenty-four patients with brain tumors were imaged in this retrospective study. There were 13 males and 11 females, with average age of 51±15 years. For therapeutic monitoring, eight patients with grade IV glioblastoma were imaged for assessment of treatment response and tumor progression. Histological and genomic profile markers were recorded. The molecular markers were isocitrate dehydrogenase gene (IDH1) mutation status, ATP-dependent helicase (ATRX), TP53, 1p/19q co-deletion, O6-methylguanine-DNA methyltransferase (MGMT) methylation status.

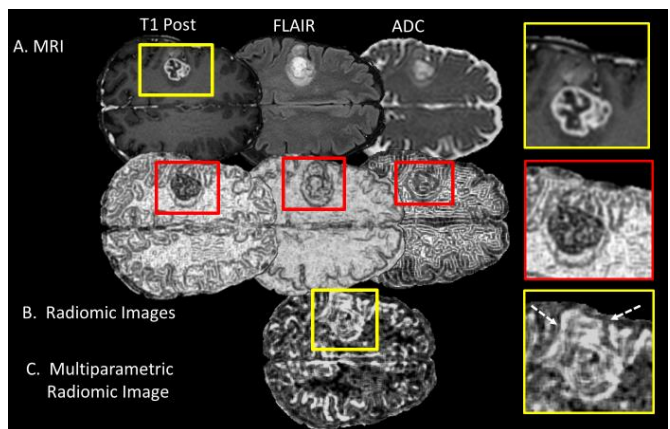

Fig. 1. A) Example multiparametric MRI from a Grade IV tumor with the post contrast enhanced T1 image displaying the perirolandic lesion (Yellow Box). B) Single radiomic images from each of the MRI parameters. C) Multiparametric radiomics of all the MRI data together demonstrating improved tissue characteristics.

**MRI Sequences:** MR images were obtained using a 3.0 Tesla Siemens Trio Tim system (Siemens Medical Solutions, Erlangen, Germany) with transmit/receive head matrix coil.

**Anatomical Sequences:** Structural images included a 3D T1 MPRAGE sequence (TR = 2300ms, TI = 900ms, TE = 3.5ms, flip angle = 9°, field of view = 24cm, acquisition matrix = 256 x 256 x 176, slice thickness (ST) = 1mm) and a 2D T2-FLAIR axial sequence (TR=9310 ms, TI=2500 ms, TE=116 ms, flip angle = 141°, field of view = 24cm, acquisition matrix = 320 x 240 x 50, ST= 3 mm).

**Perfusion and Diffusion Sequences:** Perfusion Weighted Imaging: Whole brain single-shot GRE EPI sequence after administration of 0.1mmol/kg of a gadolinium-based contrast agent was acquired with parameters as follow: TR/TE=2450/45ms; 90º deg flip angle; FOV=240x240mm 128x128 matrix; ST=4mm with 1 mm interslice gap; 32 volumes acquired with first 2 discarded to allow MR signal to reach steady state.

Diffusion Weighted Imaging: Diffusion tensor imaging was obtained using spin echo EPI parallel imaging sequence TR/TE=6700/90, 90º flip angle, 192x192mm FOV, 96x 96 matrix, 3mm, 20 directions of diffusion encoding, and b values = 0 and 1000 s/mm2. Trace monoexponentially Apparent Diffusion Coefficient (ADC) of water maps were constructed from the DWI.

**Registration methods**: For accurate co-localization, we have developed a hybrid registration method based on a combined 3D wavelet transformation with nonlinear affine transformation that performs 3D resampling and interpolations of the reference and target radiological images without a loss of information [24]. The hybrid registration scheme consists of reslicing and matching each modality using a combination of wavelet and affine registration steps. First, orthogonal reslicing is performed to equalize the FOV, matrix sizes, and the number of slices using a wavelet transformation between the data sets, then followed by angular resampling of the target data to match the reference data set. Finally, using the optimized angles from resampling, the registration is performed based on the similarity transformation between the reference volume and the resliced target volume. For each data set, the mean square error and Dice similarity measures were calculated.

**Segmentation methods:** We segmented the normal tissue (WM, GM and CSF) and the lesion tissue from multiparametric MRI using the multiparametric deep learning tissue signature model (MPDL) [25]. The MPDL is based on stacked sparse autoencoders (SSAE), which combines unsupervised pre-training with supervised fine tuning to counter the sparsity in the training labels. The tumor segmentation was performed in a semi-supervised fashion, where-in few pixels corresponding to every tissue type were labeled for each patient followed by subsequent training and segmentation using MPDL[26-28].

**Multiparametric Radiomics Framework:** The textural and shape properties of the brain tumor were defined by the mpRadiomics features based on tissue signature probability matrix (TSPM) and the tissue signature co-occurrence matrix (TSCM) which are basically the extensions of the first order statistics and gray level co-occurrence matrix-based features extracted from single images [19]. The input parameters for mpRadiomics analysis were determined using empirical analysis based on image resolution as follows:

   a. Binning for TSPM = 64
   b. Gray level quantization for TSCM = 64
   c. The distance d for TSCM was set to one voxel.
   d. The TSCM was evaluated in all the four directions – 0°,45°,90° and 135°.

The binning and gray level quantization for the task of distinguishing true-progression from pseudo-progression was set to 16 because of the lower image resolution of the perfusion MRI compared to the important MRI parameters (FLAIR, $T_1$-

Pre and $T_1$-Post contrast) used for distinguishing between grade II from grade IV brain tumors.

**The Radiomic Tissue Signature Model:** We define a tissue signature (TS) that represents a composite feature representation of a tissue type based on each of the different imaging sequences and demonstrated in **Fig. 1**. Mathematically, for N different imaging parameters with TS at a voxel position, Sp is defined as a vector of gray level intensity values at that voxel position, p across all the (N) images in the data sequence for different tissue types and is given by the following equation,

$$S_p = \left[I_p^{(1)}, I_p^{(2)}, I_p^{(3)}, \ldots, I_p^{(N)}\right]^T \quad (1)$$

Where, $I_p$ is the intensity at voxel position, p on each image.

**The Tissue signature probability matrix features**

The tissue signature probability matrix (TSPM) characterizes the spatial distribution of tissue signatures within a ROI. The mathematical formulation of TSPM is defined as: Suppose that the intensity values representing each voxel are quantized to some $G$ level, then the total number of possible tissue signatures in a dataset consisting of $N$ images will be equal to $G^N$. We define a function $f: T \to M$, where T is the set of all tissue signatures in the dataset and M is a N dimensional matrix with edges of length G where each tissue signature is represented as a cell. The function $f$ populates each cell of the matrix M with the frequency of occurrence of the corresponding tissue signature in the set T. The resulting matrix $M$ is called the tissue signature probability matrix (TSPM). The information content of the N dimensional multiparametric imaging dataset $(X_1, X_2, \ldots X_N)$ can be analyzed by computing the joint entropy, uniformity, and mutual information of the resultant TSPM[29]. These features are defined below.

1. The TSPM entropy, $H$ is given by the following equation:

$$H(X_1, X_2, \ldots X_N) =$$
$$-\sum_{i_1=1}^{N_g} \sum_{i_2=1}^{N_g} \ldots \sum_{i_N=1}^{N_g} TSPM(i_1, i_2, \ldots, i_N) \log_2 TSPM(i_1, i_2, \ldots, i_N)$$

2. The TSPM uniformity, $U$ is given by the following equation

$$U(X_1, X_2, \ldots X_N) = \sum_{i_1=1}^{N_g} \sum_{i_2=1}^{N_g} \ldots \sum_{i_N=1}^{N_g} TSPM(i_1, i_2, \ldots, i_N)^2 \quad (3)$$

3. The TSPM mutual information, $MI$ is given by

$$MI(X_1; X_2; \ldots; X_N) = (H(X_1) + H(X_2) + \cdots + H(X_N)) -$$
$$\cdots + \cdots (-1)^{N-1} H(X_1, X_2, \ldots, X_N) \quad (4)$$

By choosing different possible subsets $Y \subseteq \{X_1, X_2, \ldots, X_N\}$ and different values of H(Y), U(Y) and MI(Y) can be obtained producing a large number of mpRad features.

**Tissue signature first order statistics features:** The tissue signature first order statistics (TSFOS) features characterize the distribution of voxel intensities across all the imaging parameters. This is similar to a traditional first order histogram, except, the TSFOS histogram is computed from the voxel intensities across all the imaging parameters, which can be very useful when analyzing multiparametric imaging sequences such DWI and PWI. Let the Tissue Signature Histogram ($TSH$) represent a TSFOS histogram that is computed by dividing the voxel intensities in mpMRI into $B$ equally spaced bins. The first order statistical features (e.g. entropy) can be computed from the $TSH$ using the following equation:

$$Entropy_{TSFOS} = -\sum_{i=1}^{B} TSH(i) \log TSH(i) \quad (5)$$

The remaining TSFOS features, such as uniformity and energy, are similarly derived from TSH.

**Tissue signature co-occurrence matrix features**

The tissue signature co-occurrence matrix (TSCM) characterizes the spatial relationship between tissue signatures within a ROI. The TSCM is defined similar to the gray level co-occurrence matrix (GLCM) using two input parameters, distance (d) and angle ($\theta$) between two tissue signature locations [30,31]. Mathematically, the GLCM between any two tissue signatures, $S_i$ and $S_j$ is given by the following equation

$$GLCM_d^\theta(S_i, S_j, m, n) = |\{r : S_i(r) = m, S_j(r) = n\}| \; \forall \, m, n \in \{1,2,3, \ldots, G\} \quad (6)$$

where $r \in N$ (number of imaging sequences) and $|\cdot|$ denotes the cardinality of a set.

Given a distance, $d$ and angle, ($\theta$), the TSCM co-occurrence matrix for all such possible pairs of tissue signatures is given as follows:

$$TSCM_d^\theta(m, n) = \Sigma_{i,j} GLCM_d^\theta(S_i, S_j, m, n)$$
$$\forall \, i, j \; satisfied \; by \; d \; and \; \theta \quad (7)$$

Here, $TSCM_d^\theta$ is the tissue signature co-occurrence matrix. The TSCM can then be analyzed to extract twenty-two different TSCM features using the equations developed by Haralick et al[30].

**Classification:** The patients were classified using the IsoSVM algorithm [32]. The IsoSVM classification algorithm is an extension of the linear binary SVM algorithm, where in, the Isomap algorithm is pre-applied as a nonlinear kernel on the input dataset to transform the patients from a nonlinear space to a linear space. We evaluated the efficacy of the IsoSVM algorithm in classifying grade II from grade IV tumors using leave one out cross validation.

**Statistical analysis:** Summary statistics (mean and standard error of the mean) were recorded for the radiomic features. Student's t-test was performed to compare the grade II, grade IV and normal volunteers. We computed the area under the receiver operating characteristic (ROC) curve (AUC) for each of the radiomic features. Differences in the mpRadiomic features between two different time points were used to distinguish true from pseudo-progression using the Student's t-test. Statistical significance was set at p<0.05.

## III. RESULTS

**Tumor cases**: Twenty-four patients (13 males, 11 females) with brain tumors were scanned. There was no age difference (age range 27-75 years; mean 51.0, SD 15.01). Of the 24 cases, there were 9 (37.5%) Grade II and 15 (62.5%) Grade IV tumors.

**Grade II vs Grade IV**: The texture analysis of tumors of grade II and IV patients demonstrated that grade II tumors had a significantly different texture signature than grade IV tumors. For example, grade II tumors had a significantly lower contrast than grade IV tumors, owing to the presence of necrotic region in grade IV tumors (TSCM contrast G2: 4.90±0.48, G4: 15.19±0.2.18, p < 0.001).). In addition, grade II tumors were significantly more homogeneous than grade IV tumors (TSCM homogeneity G2: 0.57±0.01, G4: 0.49±0.01, p < 0.001). These features have been summarized in Table 1. Figure 2A demonstrates the TSCM mpRadiomic images on an example grade II and a grade IV patient. The IsoSVM algorithm using leave-one-out cross validation achieved a sensitivity and specificity of 93% and 100%, respectively, with an AUC of 0.95. The two-dimensional space obtained using Isomap applied to the high dimensional mpRadiomic feature space is shown in Figure 2(B) and the ROC curve for the IsoSVM classifier has been shown in Figure 2(C).

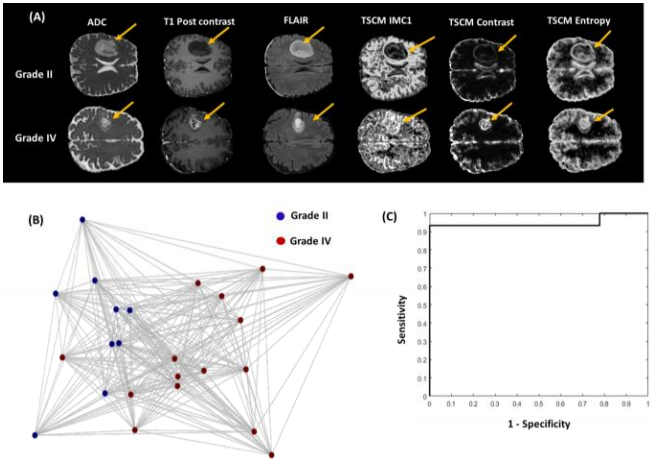

**Fig. 2** A) Example grade II and grade IV patients with the multiparametric MRI images and the corresponding multiparametric radiomics (mpRadiomic) TSCM features of informational measure of correlation 1, contrast and entropy. B) Patient scattergram demonstrating the grade II and grade IV patients in a two-dimensional space transformed by applying Isomap on the high dimensional mpRadiomic space. C) The receiver operating characteristic curve (ROC) corresponding to the IsoSVM algorithm applied on the mpRadiomic features for classification of grade II from grade IV tumors. The AUC was 0.95.

**True progression vs. Pseudoprogression:** Fig. 3 demonstrates our mpRadiomic analysis for differentiation of true progression (TP) from pseudo-progression (PP) in two representative WHO grade IV (glioblastoma) patients. There was a significant difference between the ADC map values of TP and PP. The top radiomic features of TP and PP were identified as TSCM IMC1, TSCM Correlation, TSPM entropy, and TSPM uniformity and summarized in Table 2.

The ROC curves for comparison between single and multiparametric radiomic features in distinguishing true from pseudo-progression tumors are shown Fig. 4.

## IV. DISCUSSION

The multiparametric radiomics produced excellent features for the classification of WHO grade II from grade IV brain tumors and on distinguishing true from pseudo progression after

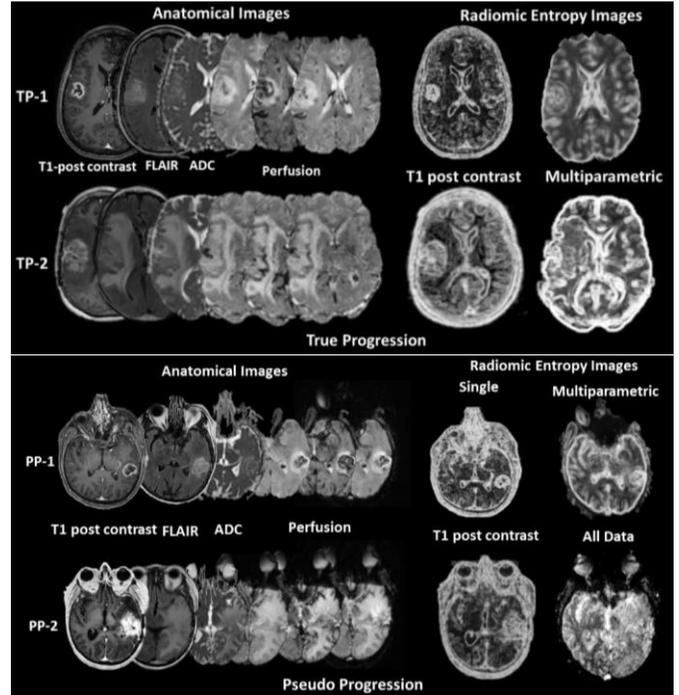

**Fig. 3** Demonstration of mpRadiomics differentiation between true from pseudo-progression of two patients with WHO Grade IV gliomas that received chemoradiation. Top Two Rows) A patient with a right temporal lobe mass at two-time points TP-1 and TP-2 demonstrating true progression. Bottom Two Rows) A patient with left temporal lobe glioblastoma at two time points with pseudo-progression (PP-1 and PP-2).

**Table 1.** Summary of the multiparametric radiomic tissue signature co-occurrence matrix (TSCM) features for classification of grade II from grade IV brain tumors.

| mpRadomic Feature | Grade II | Grade IV | p value |
|---|---|---|---|
| Contrast | 4.90±0.48 | 15.19±2.18 | 0.0003 |
| Correlation | 0.97±0.01 | 0.84±0.03 | 0.0003 |
| Entropy | 5.26±0.11 | 5.52±0.12 | 0.13 |
| Homogeneity1 | 0.57±0.01 | 0.49±0.01 | 0.00005 |
| Informational measure of correlation 1 | 0.42±0.01 | -0.28±0.02 | p<0.00001 |

chemoradiation. WHO grade IV tumors are characterized by the T1-weighted images following gadolinium administration and the presence of a necrotic core, when there is no enhancement in WHO grade II tumors. We observed that WHO grade IV tumors had mpRadiomic features of increased heterogeneity (entropy, mutual information, homogeneity) compared to grade II tumors. In addition, WHO grade IV tumors had a higher TSCM contrast than WHO grade II tumors. The higher heterogeneity and TSCM contrast could characterize the presence of more aggressive cancer in these tumors. Single image radiomics has been previously shown to characterize brain tumors in previous studies[20,21,33-35]. However, this study evaluated the efficacy of mpRadiomic features that integrates all the mpMRI parameters to produce unique radiomic features

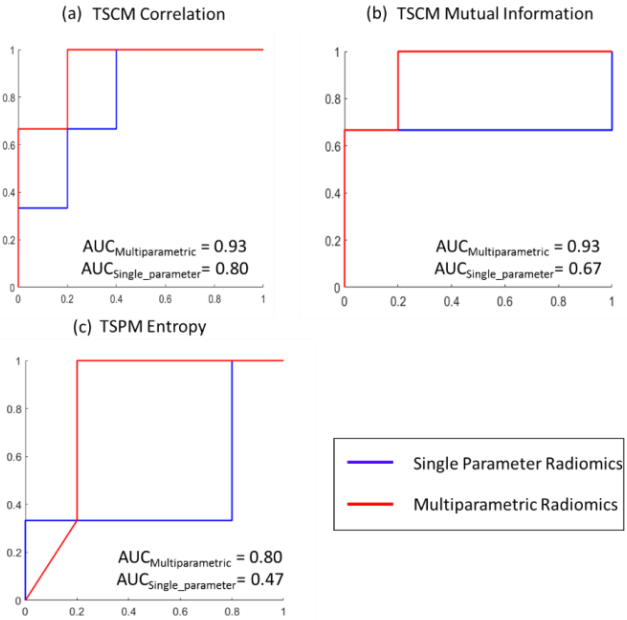

**Fig. 4** Illustration of the receiver operating characteristic (ROC) curves for classification of pseudo-progression from true-progression brain cancer for single and multiparametric radiomic features

**Table 2** Summary of the multiparametric radiomic tissue signature co-occurrence matrix (TSCM) features for classification of pseudo-progression from true-progression in glioblastomas.

| mpRadiomic Feature | Pseudo Progression | True Progression | p value | AUC |
|---|---|---|---|---|
| ADC Map (x10$^{-3}$ mm$^2$/s) | 1587±326 | 1146±193 | 0.05 | |
| Correlation | 0.88±0.02 | 0.94±0.02 | 0.11 | 0.93 |
| Informational measure of correlation 2 | 0.87±0.02 | 0.93±0.01 | 0.05 | 0.93 |
| Multidimensional entropy | 7.91±0.73 | 9.76±0.36 | 0.07 | 0.80 |
| Multidimensional uniformity | 0.0060±0.0020 | 0.0012±0.0003 | 0.08 | 0.80 |

for the prediction of brain tumor grade.

Preliminary analysis on distinguishing true from pseudo-progression demonstrated that the mpRadiomic features could provide higher classification accuracy than single parameter radiomic features. The mpRadiomic features of true progression had a different tissue texture types than the pseudo-progression cases. This distinction could be helpful in determining the appropriate therapeutic choice for the patient. However, these mpRadiomic features would need to be validated in a larger patient cohort for stability and repeatability before these features could be used as biomarkers in the clinical setting.

In general, mpMRI in brain produces a large number of images corresponding to each slice location, thereby producing a high dimensional image space. Extracting only single radiomic features from such datasets will not provide complete information about the tissue. The mpRadiomic approach resolves this issue by extracting intrinsic radiomic features of each tissue texture but also evaluates the overall tissue texture in all the images, giving a more complete picture of the tissue.

There are some limitations in using mpRadiomics in practice specific to the present study. This preliminary study was retrospective and any assessment of the clinical value of mpRadiomics will require additional prospective studies with subsequent follow-up and pathological correlation.

## V. CONCLUSION

In conclusion, the mpRadiomic analysis was able to effectively capture the multiparametric brain MRI texture and demonstrated that it potentially could be used as an imaging biomarker for distinguishing WHO grade IV from grade II tumors and true-progression from pseudo-progression after chemoradiation therapy in brain cancer.

ACKNOWLEDGEMENTS

This work was supported in part by the National Institutes of Health (NIH) grant numbers: 5P30CA006973 (Imaging Response Assessment Team - IRAT), U01CA140204, 1R01CA190299, and The Tesla K40s used for this research was donated by the NVIDIA Corporation.

**Code availability:** Our software will be freely available to academic users after issue of pending patents and a materials research agreement is obtained from the university. Due to University regulations, any patent pending software is not available until a patent is issued.
**Data availability:** All relevant clinical data are available upon request with adherence to HIPPA laws and the institutions IRB policies.
**Author Contributions:** Vishwa S. Parekh, John Laterra, Chetan Bettegowda Jay J. Pillai, Michael A. Jacobs,
**Conflict of Interest:** The authors have no conflict of interests.